\documentclass[superscriptaddress,aps,reprint,onecolumn,notitlepage]{revtex4-1}
\pdfoutput=1
\usepackage{amssymb,amsmath}
\usepackage{xcolor}
\usepackage{graphicx}
\usepackage{ulem}
 \usepackage[colorlinks=true,
 linkcolor=black,
 urlcolor=blue,
 citecolor=blue]{hyperref}

\newcommand\numberthis{\addtocounter{equation}{1}\tag{\theequation}}

\newcommand{\comments}[1]{}

\definecolor{ao}{rgb}{0.0, 0.4, 0.0}

\def\d{\mathrm{d}}

\begin{document}
\title{Phenotypic switching of populations of cells in a stochastic environment}
\author{Peter G. Hufton}\email{peter.hufton@postgrad.manchester.ac.uk}
\affiliation{Theoretical Physics, School of Physics and Astronomy, The University
of Manchester, Manchester M13 9PL, United Kingdom}

\author{Yen Ting Lin}
\email{yentingl@lanl.gov}
\affiliation{Theoretical Physics, School of Physics and Astronomy, The University
of Manchester, Manchester M13 9PL, United Kingdom}
\affiliation{Theoretical Division and Center for Nonlinear Studies, Los Alamos National Laboratory, Los Alamos, New Mexico 87544, USA}
\author{Tobias Galla}\email{tobias.galla@manchester.ac.uk}
\affiliation{Theoretical Physics, School of Physics and Astronomy, The University
of Manchester, Manchester M13 9PL, United Kingdom}

\begin{abstract}
In biology phenotypic switching is a common bet-hedging strategy in the face of uncertain environmental conditions. Existing mathematical models often focus on periodically changing environments to determine the optimal phenotypic response. We focus on the case in which the environment switches randomly between discrete states. Starting from an individual-based model we derive stochastic differential equations to describe the dynamics, and obtain analytical expressions for the mean instantaneous growth rates based on the theory of piecewise-deterministic Markov processes. We show that optimal phenotypic responses are non-trivial for slow and intermediate environmental processes, and systematically compare the cases of periodic and random environments. The best response to random switching is more likely to be heterogeneity than in the case of deterministic periodic environments, net growth rates tend to be higher under stochastic environmental dynamics. The combined system of environment and population of cells can be interpreted as host-pathogen interaction, in which the host tries to choose environmental switching so as to minimise growth of the pathogen, and in which the pathogen employs a phenotypic switching optimised to increase its growth rate. We discuss the existence of Nash-like mutual best-response scenarios for such host-pathogen games.
\end{abstract}

\maketitle

\section{Introduction}
\label{sec:introduction}

Outside of laboratory conditions, microbial cells are often subject to unpredictable adverse environmental changes. As a mechanism to survive such changes, we now understand that cells have evolved to switch stochastically between phenotypic states \cite{wiuff2005phenotypic,levin2006non,smits2006phenotypic,leisner2008stochastic,choi2008stochastic}. In contrast to more familiar forms of resistance caused by permanent genetic mutations, these switches are epigenetic in nature, reversible, and lead to heterogeneity in the population.
Microbial cells have been shown to use this mechanism as a survival strategy to react to changing environments, for example induced by the administration of antibiotics or other short-term changes in environmental conditions \cite{lewis2007persister,balaban2004bacterial}. Due to the continued interest in antibiotic resistance, understanding the use of phenotypic switching as a survival strategy is essential.

This paper focuses on modelling the dynamics of phenotypic switching in bacterial populations. Within such a population, variability from cell to cell can lead to an fitness advantage. A classic study by Bigger \cite{bigger1944treatment} suggests the existence of a subpopulation of bacterial more resistant to antibiotics than other members of the population; these more resilient cells are often referred to as `persisters' \cite{dhar2007microbial,gefen2009importance,rainey2011evolutionary}. The difference between `normal' and `persister' cells is phenotypic rather than genotypic. Consequently, a cell may switch back and forth between the normal and persister states, and a given population will contain subpopulations of both phenotypes simultaneously. This heterogeneity in the population constitutes a strategy to improve resistance to environmental stresses, a dynamic sometimes described as bet-hedging \cite{grafen1999formal}. Biological systems use this form of risk-balancing against a number of different environmental stresses, including differences in temperature, concentrations of nutrients and toxins, or the state of a host immune response \cite{lewis2007persister,acar2008stochastic,hallet2001playing,tuchscherr2011staphylococcus}; Thattai {\it et al.} \cite{thattai2004stochastic}, for example, suggest a model considering the growth of cells in the urinary tract where the environment relates to the state of urination. The effects of environmental sources of noise on biochemical networks has been an important area of theoretical research, where it has been shown to modify switching rates \cite{shahrezaei2008colored,hu2011effects,leisner2009kinetics} and induce bistability \cite{samoilov2005stochastic,caravagna2013interplay}. Mathematical models of such bet-hedging strategies often focus on periodically varying environments, mainly to keep the analysis manageable. Examples can be found in \cite{patra2015emergence,belete2015optimality,lachmann1996inheritance,balaban2004bacterial,kussell2005bacterial,acar2008stochastic,leibler2010individual,gaal2010exact,lohmar2011switching,patra2013population}. To a lesser extent, bet-hedging has also been studied in stochastically varying environments \cite{levins1968evolution, haccou1995optimal,thattai2004stochastic,kussell2005phenotypic,gander2007stochastic,visco2010switching,horvath2016study}. 

In this paper, we present a mathematical framework to model phenotypic switching as a mechanism of persistence in a fluctuating environment. 
We begin by considering an individual-based model, where possible `reactions' include random birth and death events and stochastic phenotypic switching of individual cells, both in a periodically and stochastically changing environments.
We use a recently-developed mathematical approach \cite{hufton2016intrinsic} to characterise the joint random processes and derive formulae for the fitnesses of the subpopulations and the overall growth rate. Further analysis shows that heterogeneity is advantageous for slowly switching environments, whereas homogeneity conveys a fitness advantage for quickly switching environments. By comparing our results to the case of a periodic environment, we find that environmental stochasticity can, in principle, lead to strategies of phenotypic switching which are very different from those in periodic external environments.
 
The implications of our analysis are twofold. Firstly, the method advances the mathematical framework for analysing ecological models in random environments. It provides not only a more efficient computational scheme for studying growth dynamics than direct simulations, but also offers mechanistic insights into the problem. Secondly, the method can be used to investigate effective protocols of administering treatment, such as antibiotics to bacterial infection in a control-theoretic framework. We demonstrate both of these applications in idealised scenarios in the later sections of our paper. We also identify a scenario where the competition between the switching dynamics of a host environment and the phenotypic response of the pathogen constitute a game-theoretic scenario. Our analysis allows us to identify the optimal control strategy for the host and the optimal survival strategy for the pathogen. We show how this can result in mutual best responses, akin to what is known as Nash equilibria in game theory \cite{nash1951non}.

The remainder of the paper is organised as follows. In Sec.~\ref{sec:methods} we set up an individual-based model of phenotypic heterogeneity in a stochastic environment. We then approximate the dynamics of this model and obtain a solution for the average growth rate of the population of cells (Sec. \ref{sec:analysis}). Using this solution, we analyse how the average growth rate changes with the parameters in Sec.~\ref{sec:results}. Specifically, we look into the optimum bet-hedging strategy for the bacteria to maximise growth rate, and conversely the environmental switching strategy that best hinders the bacteria's growth rate. In Sec.~\ref{sec:conclusion} we summarise our results, and discuss possible directions for future work.

\begin{figure*}
\includegraphics[width=0.90\textwidth]{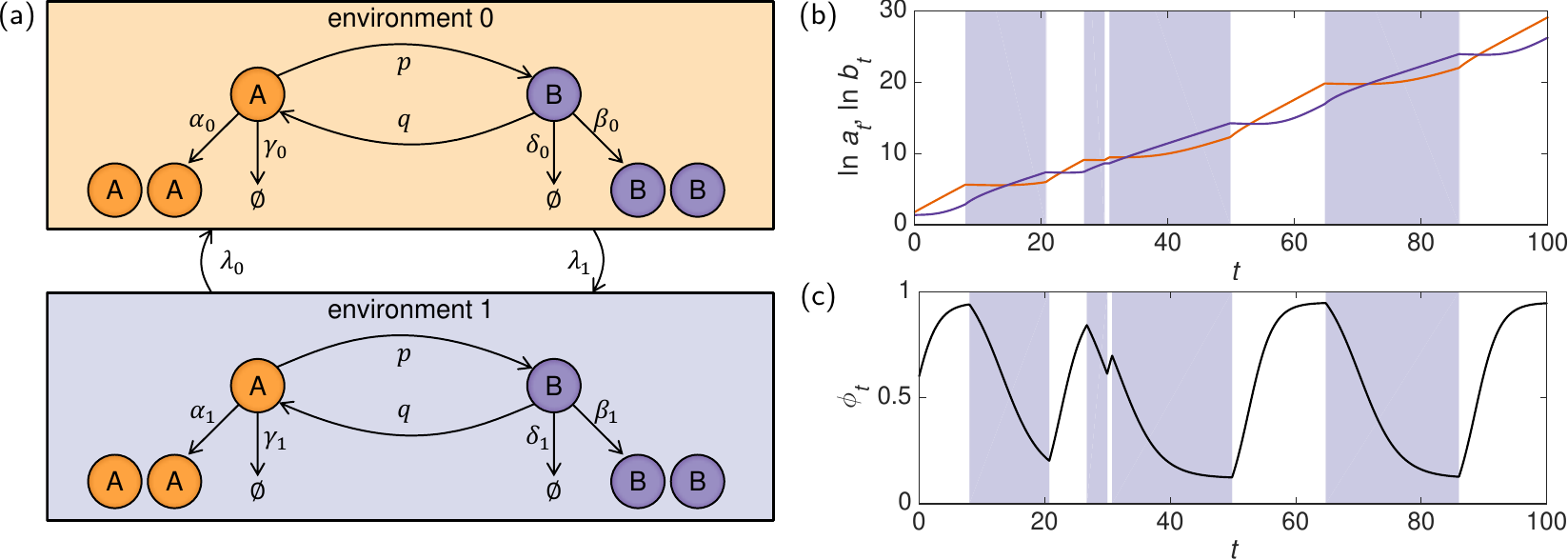}
\caption{(a) Schematic overview of the model. Each cell can express either phenotype $A$ or phenotype $B$, and switches from $A$ to $B$ with rate $p$, and vice versa with rate $q$. Cells of type $A$ duplicate with rate $\alpha_\sigma$ and die with rate $\gamma_\sigma$, where $\sigma\in\{0,1\}$ is the state of the environment. The rates for phenotype $B$ are $\beta_\sigma$ and $\delta_\sigma$. The environment switches from state $0$ to state $1$ with rate $\lambda_1$, and vice versa with rate $\lambda_0$. The rates are such that the growth rate in environmental state $0$ is higher for cells with phenotype $A$, and the growth rate in environmental state $1$ is higher for cells with phenotype $B$ ($\mu_0^A>\mu_0^B$, $\mu_1^A<\mu_1^B$).
(b) A typical trajectory of the populations as a piecewise-deterministic Markov process (see text for details, phenotype $A$ orange, phenotype $B$ purple). The background shading indicates whether the environment is in state $0$ (light) or state $1$ (dark). The population of phenotype $A$ is typically greater than that of phenotype $B$ in environment $0$ and vice versa in environment $1$.
(c) The corresponding trajectory of the proportion of the population expressing phenotype $A$. As the environment switches states the proportion of phenotype $A$ alternately tends towards the two fixed points, $\phi_0^+$ and $\phi_1^+$, discussed in the text.
Simulations use parameter set (b) (see text and Appendix~\ref{sec:parameters}). Environmental switching rates are $\lambda_1=0.10$, $\lambda_1=0.10$ and phenotypic switching rates are $p=0.028$, $q=0.043$.}
\label{fig:schematic}
\end{figure*}

\section{Model definitions}
\label{sec:methods}
\subsection{Individual-based model} We study an individual-based model of a well-mixed population of a species with phenotypic switching in a changing environment. 
Figure~\ref{fig:schematic}(a) presents a summary of the model. We consider a simplified scenario of two phenotypes and two environmental states. Each cell in the population expresses either phenotype $A$ or phenotype $B$, representing the `normal' and `persister' phenotypes, respectively. The birth and death dynamics of the phenotypes depend on an environmental state, which we label $\sigma$ and which takes values $0$ or $1$. State $0$ represents the `growth state'; in this state external stresses on the population are absent and the normal phenotype outperforms the persister phenotype. State $1$ on the other hand, is the `stress state' and indicates that an external stress acts on the population. In this state, the persister phenotype outperforms the normal phenotype.
This stress can for example be the presence of antibacterial agents, the absence of nutrients, unfavourable temperatures, or the state a host immune response.

In our model individuals expressing phenotype $A$ duplicate with rate $\alpha_\sigma$ and die with rate $\gamma_\sigma$. The corresponding rates for phenotype $B$ are $\beta_\sigma$ and $\delta_\sigma$. We define the net growth rates $\mu_\sigma^A=\alpha_\sigma-\gamma_\sigma$ and $\mu_\sigma^B=\beta_\sigma-\delta_\sigma$. Cells expressing phenotype $A$ are better suited to environment $0$, whereas cells expressing phenotype $B$ are better suited to environment $1$. Specifically, parameters are chosen such that $\mu_0^A>\mu_0^B$ and $\mu_1^A<\mu_1^B$. Individuals switch between the two phenotypes stochastically \cite{choi2008stochastic}: individuals of phenotype $A$ switch to phenotype $B$ with rate $p$, switches in the opposite direction occur with rate $q$. 
The rates $p$ and $q$ constitute the `switching strategy' of the population.
These rates are assumed to be independent of the environment, i.e., cells switch between types $A$ and $B$ without sensing the state of the environment.
The mathematical formalism we develop, however, does not rely on this assumption and can readily be extended to include environmental sensing, as considered in Refs.~\cite{thattai2004stochastic,kussell2005phenotypic,donaldson2010fitness,rivoire2011value,pugatch2013asymptotic,ogura2017delayed}. More specifically, all equations in this section can be modified to consider environmental sensing under the replacement of phenotypic switching rates $p$ and $q$ with environment-dependent quantities $p_\sigma$ and $q_\sigma$.
 
We focus mainly on the case when the environmental switching follows a Markov process, i.e., the environmental process is memoryless. We also assume that the dynamics of the environment are external, in particular it is independent of the population. Switches from state $\sigma=0$ to state $\sigma=1$ occur with constant rate $\lambda_1$, and switches from $1$ to $0$ occur with constant rate $\lambda_0$. As a consequence, the lengths of the episodes spent in either environment are identically and independently distributed random variable sampled from exponential distributions with rate parameters $\lambda_0$ and $\lambda_1$, respectively. Mathematically, the environment is described by an asymmetric random telegraph process \cite{gardiner1985handbook}.

In real-world systems the environmental switching process will often have memory and will not occur at constant rates. Instead, transitions of the environmental state will be more likely to occur at certain moments in time. In the extreme case of a laboratory setting they may be strictly prescribed by a definite experimental protocol. The case of strictly periodic switching between environmental states is regularly assumed in the literature, mostly for mathematical convenience. But, just like the Markovian case this not entirely realistic either. The main motivation for comparing the cases of a Markovian environment on the one hand, and a strictly periodic environment on the other, is that these cases serve as two extremes; they delimit the regime of real-world controlled environmental switching statistics, and by studying the two extremes, we shed light on more realistic scenarios operating in-between. 

In principle, the model described above could be simulated using the Gillespie algorithm \cite{gillespie1977exact}. In practice, this method is slow, in particular if the total population grows with time. Instead we therefore use a set of approximations which allow us to simulate the processes more efficiently, and also to obtain analytical results.
Starting from the master equation for the model, we approximate the dynamics in the limit of a large but finite system size. This leads to a diffusion process with Markovian switching \cite{mao2006stochastic}, and allows us to perform the so-called linear-noise approximation \cite{van1992stochastic}. From this we obtain explicit expressions for the average growth rate of the population.

\subsection{Diffusion process with Markovian switching and linear-noise approximation.}
Since our model is Markovian, the evolution of the probabilities of the system to be in a specific state is described by a master equation \cite{van1992stochastic,gardiner1985handbook,risken1984fokker}. 
Except for a few special cases, however, there exists no analytical solution for a general master equation. We build on recent work \cite{hufton2016intrinsic, lin2016gene, lin2016bursting} to approximate the dynamics of systems with switching environments, and perform a Kramers--Moyal expansion \cite{van1992stochastic,risken1984fokker,gardiner1985handbook} in the limit of a large population of cells, but maintaining the discreteness of the environment. Further details can be found in Appendix~\ref{sec:derivation}. The result of this expansion for our model is a set of two coupled Ito stochastic differential equations (SDEs) for the number of individuals $a_t$ and $b_t$ with phenotypes $A$ and $B$ respectively. These equations are
\begin{subequations}\begin{align}
\begin{split}\d a_t=&\left(\mu_\sigma^A a_t -pa_t+qb_t\right)\d t + {B_\sigma}_{11}(a_t,b_t) \d W_t^{(1)} + {B_\sigma}_{12}(a_t,b_t) \d W_t^{(2)},\label{eq:SDE_a}\end{split}\\
\begin{split}\d b_t=&\left(\mu_\sigma^B b_t +pa_t-qb_t\right)\d t + {B_\sigma}_{21}(a_t,b_t) \d W_t^{(1)} +{B_\sigma}_{22}(a_t,b_t) \d W_t^{(2)}.\label{eq:SDE_b}\end{split}
\end{align}\label{eq:SDE_a_b}\end{subequations} 
In the limit of large populations, the formerly-discrete numbers of individuals $a_t$ and $b_t$ are approximated as continuous variables in Eqs.~\eqref{eq:SDE_a_b}. The quantities $W^{(1)}_t$ and $W^{(2)}_t$ are independent Wiener processes (Brownian motion). The subscript $t$ is used throughout this paper to indicate a random process.
The coefficients ${B_\sigma}_{ij}(a,b)$ characterise the strength of the demographic noise arising from the random birth-and-death process and the phenotypic switching events \cite{van1992stochastic,gardiner1985handbook}. 
We provide the exact forms of ${B_\sigma}_{ij}$ in Appendix~\ref{sec:derivation}. 

Equations~\eqref{eq:SDE_a_b} describe the random evolution of the population of bacteria {\em between} switching events of the environment \cite{mao2006stochastic}. In particular the quantities on the right-hand side with a subscript $\sigma$ depend on the current state of the environment. For Markovian switching of the environment as described above, the environmental process $\sigma_t$ is governed by the master equation
\begin{equation}
\frac{\d}{\d t}P_{\sigma}(t)=\lambda_\sigma P_{1-\sigma}(t) - \lambda_{1-\sigma} P_{\sigma}(t),
 \label{eq:environ_CME}
\end{equation}
where $P_{\sigma}(t)$ denotes the probability that the environment is in state $\sigma\in\{0,1\}$ at time $t$.

In our analysis, we also consider a periodic environment. In this case the state of the environment is a deterministic function given by
\begin{equation}
\sigma(t) = 
\begin{cases}
0 \quad\text{ for }\quad \,\,\,\, 0 \leq t' < \tfrac{1}{\lambda_1},\\
1 \quad\text{ for }\quad \tfrac{1}{\lambda_1} \leq t' < \tfrac{1}{\lambda_1}+\tfrac{1}{\lambda_0},\\
\end{cases}
\label{eq:environ_periodic}
\end{equation}
and then repeated with period $T= 1/\lambda_0+1/\lambda_1$. This is chosen such that the duration of episodes spent in environments $0$ and $1$ respectively in the periodic case match the mean episode durations of the random environmental process.

The above equations provide us with a computational scheme for simulating trajectories of the system. For stochastically switching environments, environmental switching times are generated randomly from an exponential distribution, using rate parameters $\lambda_1$ and $\lambda_0$ in environments $0$ and $1$, respectively.

Equations~\eqref{eq:SDE_a_b} are numerically integrated using the current $\sigma$ until the next switching time is reached. When the prescribed switching is reached, we switch environmental state and repeat the process. This procedure is more efficient than direct simulations of the individual-based system, especially for large populations. In particular the time to simulate Eqs.~\eqref{eq:SDE_a_b} does not increase with increasing population size, in contrast to the computing time required by the Gillespie algorithm.

By virtue of the central limit theorem, the strength of the demographic noise scales with the inverse square root of the total size of the population. Hence in the limit of large populations, the contribution of the diffusive terms in the SDEs [Eqs.~\eqref{eq:SDE_a_b}] becomes negligible. Ignoring the diffusive terms leads to a so-called piecewise-deterministic Markov process (PDMP) \cite{davis1984piecewise,faggionato2009non},
\begin{subequations}\begin{align}
\d a_t=&\left(\mu_\sigma^A a_t -pa_t+qb_t\right)\d t ,\label{eq:PDMP_a2}\\
\d b_t=&\left(\mu_\sigma^B b_t +pa_t-qb_t\right)\d t. \label{eq:PDMP_b2}
\end{align}\label{eq:PDMP_a_b}\end{subequations} 
This process evolves deterministically between stochastic environmental switching events, which occur at discrete times. This simplification allows for analytical solutions, and much of our further analysis will be carried out in this limit.

Figure \ref{fig:schematic}(b) shows a typical trajectory for both phenotypes and the environment using the PDMP method in a stochastically switching environment. The size of both populations increases approximately exponentially with time. This provides an \textit{a posteriori} justification for our approximation to neglect intrinsic demographic noise. We note that the population of phenotype $A$ tends to be larger in environmental state $0$, and in state $1$ phenotype $B$ is usually more abundant, at least once enough time has passed since the last environmental switch. 
\section{Analysis}\label{sec:analysis}
 
\subsection{Stochastic differential equations for population size and composition} 
The model above is deceptively simple. By ignoring demographic noise, the population dynamics of Eq.~\eqref{eq:PDMP_a_b} is a linear dynamical system, described by two coupled ordinary differential equations with Markovian switching. The standard technique to solve the dynamics of coupled linear ODEs is to find the eigenvalues and eigenvectors of the propagating matrix. Along the eigenvectors, growth or decay is exponential with rates equal to the corresponding eigenvalues. There are two subtleties in our particular case. First, when we have Markovian environmental switching, the time between environmental switches is random, and so the matrix describing the propagation of the population from the start of an episode to the end is also random. Second, the eigenvectors of the propagating matrix in one environment do generally not align with those in the other environmental state. As a result, the propagation of the population in the long run is described by products of a sequence of non-commuting random matrices. This makes further analysis difficult. Progress can however be made by introducing the processes 
\begin{subequations}\begin{align}
n_t={}&a_t+b_t,\\
\phi_t={}&\frac{a_t}{a_t+b_t}.
\end{align}\end{subequations} 
They describe the total number of individuals in the population, and the proportion of individuals of phenotype $A$, respectively. 
Together, the processes $n_t$ and $\phi_t$ provide an alternative coordinate system for describing the state of the system; such coordinates have been used previously to describe competition in growing populations \cite{melbinger2010evolutionary,cremer2011evolutionary}. With these definitions, Eqs.~\eqref{eq:SDE_a}~and~\eqref{eq:SDE_b} become
\begin{subequations}\begin{align}
\begin{split}
\d n_t=&n_t\left[\mu_\sigma^A \phi_t + \mu_\sigma^B (1-\phi_t)\right]\d t
+n_t^{+\frac{1}{2}}{C_\sigma}_{11}(\phi_t) \d W_t^{(1)} + n_t^{+\frac{1}{2}}{C_\sigma}_{12}(\phi_t) \d W_t^{(2)},\label{eq:PDMP_N}
\end{split}\\
\begin{split}
\d \phi_t=&\left[\Delta_\sigma \phi_t(1-\phi_t) -p\phi_t+q(1-\phi_t) \right]\d t
+n_t^{-\frac{1}{2}}{C_\sigma}_{21}(\phi_t) \d W_t^{(1)} + n_t^{-\frac{1}{2}}{C_\sigma}_{22}(\phi_t) \d W_t^{(2)},\label{eq:PDMP_phi}
\end{split}
\end{align}\label{eq:PDMP_N_phi}\end{subequations} 
where $\Delta_\sigma=\mu_\sigma^A-\mu_\sigma^B$. 
Details of the coefficients ${C_\sigma}_{ij}(\phi)$ are given in the Appendix~\ref{sec:derivation}.

Equation~\eqref{eq:PDMP_N} describes the growth of the total population; the average per capita growth rate is given by the expression in the square bracket. Equation~\eqref{eq:PDMP_phi} describes the evolution of the proportion of cells expressing phenotype $A$.

Suppressing again the demographic noise leads to a PDMP description of $n_t$ and $\phi_t$, valid in the limit of a large system. The corresponding dynamics are given by Eqs.~\eqref{eq:PDMP_N_phi}, with the noise terms $\d W_t^{(1)}$ and $\d W_t^{(2)}$ removed. Interestingly, in this limit the evolution of the fraction $\phi_t$ decouples from the evolution of the total population: the proportion of each phenotype in the population is independent of the size of the total population and follows the nonlinear logistic equation obtained from Eq.~\eqref{eq:PDMP_phi}. It is important to stress that the decoupling of the two processes is a consequence of the linearity of the model. More general dynamics will not have this property---for these models the dynamics of $\phi_t$ and the growth rate may depend on the population size $n_t$.

Nevertheless, our approach may still offer insight into cases with non-linear interactions. A standard technique for analysing competition in laboratory experiments involves successive dilution steps of a population, in order to limit the population size to some region of biological interest and experimental practicability \cite{lechner2012staphylococcus,wiser2013long}. The dynamics and growth rates are thus considered in this limited region of $n_t$ only. For our mathematical framework we propose an analogous technique as a potential area of future research, namely to fix the population size at some specific $n$, and then to study the effective growth rate---the contents of the square bracket in Eq.~\eqref{eq:PDMP_N}. This technique may also be compared to a birth-death model such as the Moran process, in which the population size is fixed through paired birth and death events \cite{moran1962statistical}.

\subsection{Calculation of average growth rate}
The right-hand side of Eq.~\eqref{eq:PDMP_N} shows that the instantaneous growth rate is given by
\begin{equation}
\mu_t\equiv\mu_\sigma^A \phi_t + \mu_\sigma^B (1-\phi_t).
\label{eq:mu}
\end{equation}
We are interested in computing the average growth rate of the system, which we write as $\rm{E}\left(\mu\right)$. This quantity can be understood as the instantaneous growth rate $\dot n/n$, averaged over many realisations of the random processes (after a transient time has passed). Since the process $\phi_t$ is ergodic, this ensemble average is equivalent to an average over a long time. Furthermore, it is easy to show that this average instantaneous growth rate is equivalent to the average rate of the exponential growth of the population over a long time, the so-called `dominant Lyapunov exponent' \cite{metz1992should}.
To proceed, it is useful to first consider the average growth rate in a given environmental state $\sigma$, for which we write $\rm{E}\left(\mu|\sigma\right)$. This is the instantaneous growth rate averaged over all points in time at which the environment is in state $\sigma$.
In order to compute this object and given Eq.~\eqref{eq:mu} it is sufficient to know $\rm{E}\left(\phi\middle|\sigma\right)$---the average value of $\phi$ in environmental state $\sigma$. 
In order to obtain this object, an average over intrinsic and environmental fluctuations is required. We note that these intrinsic fluctuations are represented in the stochastic differential equations Eqs.~\eqref{eq:PDMP_N_phi} by multiplicative noise. Further simplification can be achieved in the linear-noise approximation (LNA) \cite{van1992stochastic,gardiner1985handbook}, valid also for large but finite populations. The LNA consists in replacing $\phi_t$ in the noise correlators (the coefficients ${C_{\sigma}}_{ij}$) by the trajectory of the PDMP. In this limit, all fluctuations about the PDMP trajectory due to intrinsic noise will then be symmetrical, and so performing the average of $\phi$ with respect to intrinsic fluctuations is equivalent to considering the PDMP process only, and to compute $E(\phi|\sigma)$ as an average over the environmental process (see also Ref.~\cite{bayati2016deterministic}). Thus, for the remainder of the analysis we concentrate on the PDMP description of $\phi_t$,
\begin{equation}
\d \phi_t=\left[\Delta_\sigma \phi_t(1-\phi_t) -p\phi_t+q(1-\phi_t) \right]\d t. \label{eq:phiPDMP}
\end{equation}

Figure~\ref{fig:schematic}(c) shows a sample path of the PDMP for the proportion of the population with phenotype $A$, $\phi_t$. The process tends towards two different fixed points depending on environmental state. An analysis of Eq.~\eqref{eq:phiPDMP} shows that there are two fixed points for each environmental state: one stable $\phi_\sigma^+$, and one unstable $\phi_\sigma^-$. These are given by
\begin{equation}
\phi_\sigma^\pm = \frac{\Delta_\sigma - p - q \pm \sqrt{(\Delta_\sigma-p-q)^2 +4q\Delta_\sigma}}{2\Delta_\sigma}.
\numberthis
\label{eq:stat}
\end{equation}
For our model, $\phi_1^+<\phi_0^+$ since state $0$ favours phenotype $A$ and state $1$ favours phenotype $B$. As the environment switches between states, the process $\phi_t$ alternates between tending towards these two stable fixed points. After sufficient time has passed, the process will be confined to the interval between the two stable fixed points $(\phi_1^+,\phi_0^+)$. 

Equation~\eqref{eq:phiPDMP} describes a single-variable PDMP in a Markovian, two-state environment. Processes of this form have been the subject of recent studies, particularly in the context of gene regulatory networks
 \cite{crudu2009hybrid,zeiser2010autocatalytic,lin2016gene, hufton2016intrinsic}. Reference~\cite{hufton2016intrinsic} provides a general solution for the stationary probability density distribution $\Pi^*_\sigma(\phi)$ for such a process. In the context of the present model, we find
\begin{subequations}\begin{align}
\begin{split}\Pi^*_0(\phi)={}\tfrac{+\mathcal{N}}{\Delta_0}
&\left( \phi_0^+-\phi \right)^{ g-1}
\left( \phi-\phi_0^- \right)^{-g-1}\left( \phi-\phi_1^+ \right)^{ h}
\left( \phi_1^- - \phi\right)^{-h},\end{split}
\\
\begin{split}\Pi^*_1(\phi)={}\tfrac{-\mathcal{N}}{\Delta_1}
&\left( \phi_0^+-\phi \right)^{ g}
\left( \phi-\phi_0^- \right)^{-g}\left( \phi-\phi_1^+ \right)^{h-1}
\left( \phi_1^--\phi \right)^{-h-1},\end{split}
\end{align}\label{eq:pdmp_dist}\end{subequations}
for $\phi\in(\phi_1^+,\phi_0^+)$. These densities are to be be interpreted as follows: $\Pi^*_0(u)\d u$ is the probability to find the system in environmental state 0 and with $\phi\in(u,u+\d u]$ in the stationary state. A similar interpretation applies to $\Pi_1(\phi)$. The exponents $g$ and $h$ are given by
\begin{align}
g=\frac{\lambda_1}{\Delta_0(\phi_0^+-\phi_0^-)}, &&
h=\frac{\lambda_0}{\Delta_1(\phi_1^+-\phi_1^-)},
\end{align}
and $\mathcal{N}$ is determined by normalisation, $\int_{\phi_1^+}^{\phi_0^+} \left[ \Pi^*_0(\phi) + \Pi^*_1(\phi) \right] \d \phi = 1$.

\begin{figure*}
\includegraphics[width=0.95\textwidth]{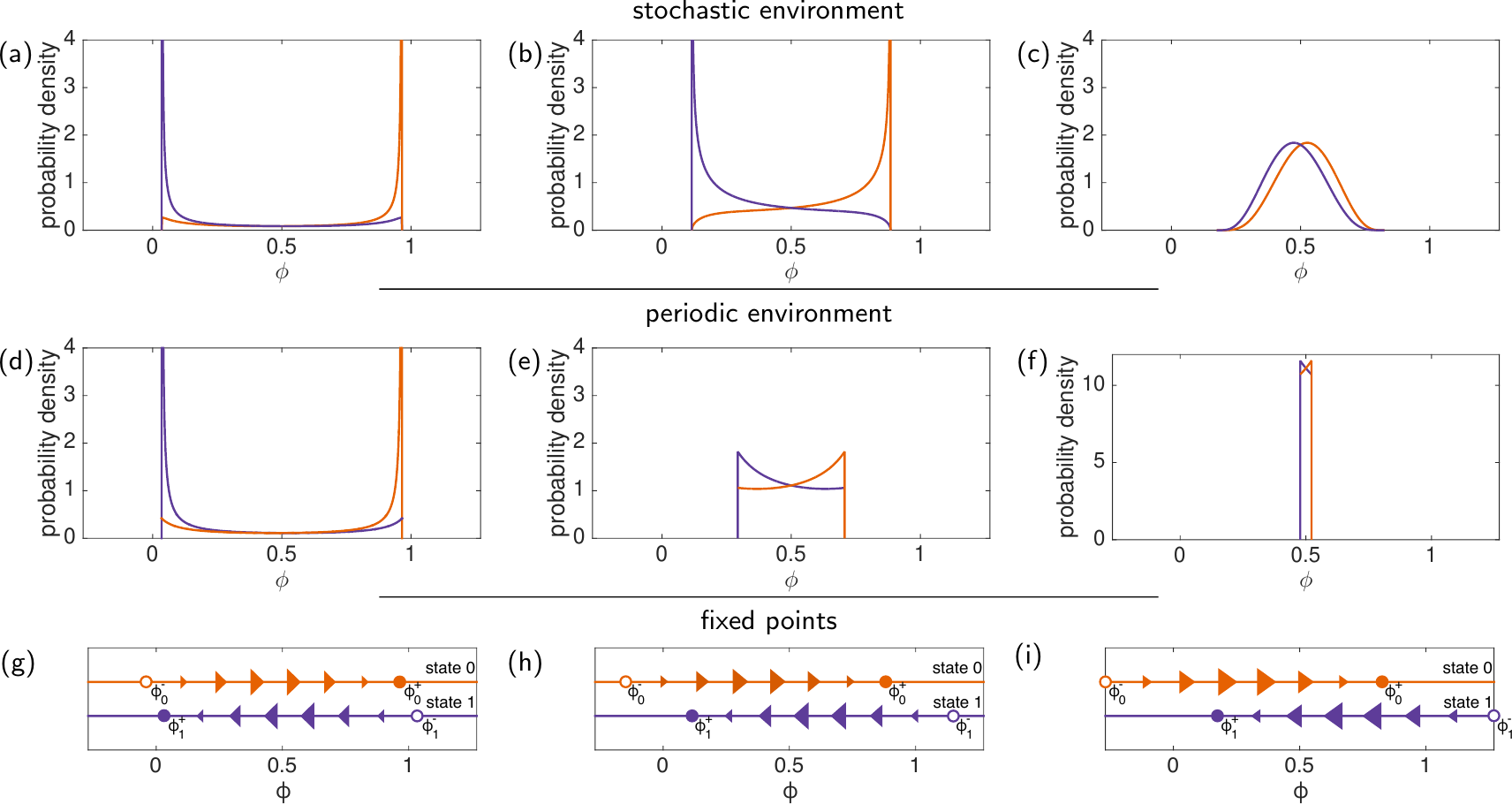}
\caption{Panels (a), (b), and (c) show the stationary distributions of $\phi$---the proportion of cell in phenotype A---for the case of a stochastic environment. These distributions are obtained from Eq.~\eqref{eq:pdmp_dist}. From left to right, panels correspond to increasing environmental switching rates. The orange line shows $\Pi^*_0(\phi)$, and the purple line $\Pi^*_1(\phi)$ (see text for details). In the case of slow environment switching (a), the distribution is sharply peaked at the two fixed points whereas in the case of fast environmental switching (c) the distribution is peaked in the central region between the two fixed points. Panels (d), (e), and (f) show the stationary distributions for the case of a periodic environment. Panels (g), (h), and (i) depict the positions of stable and unstable fixed points and the direction of flow in each state. Parameters are from set (a) (see text and Appendix~\ref{sec:parameters}), along with:
(a),(d), and (g) $\lambda_1, \lambda_0=0.10$, $p, q=0.064$;
(b),(e), and (h) $\lambda_1, \lambda_0=1.0$, $p, q=0.24$;
(c),(f), and (i) $\lambda_1, \lambda_0=10$, $p, q=0.40$.}
\label{fig:dist_phi}
\end{figure*}
\begin{figure*}
\includegraphics[width=0.95\textwidth]{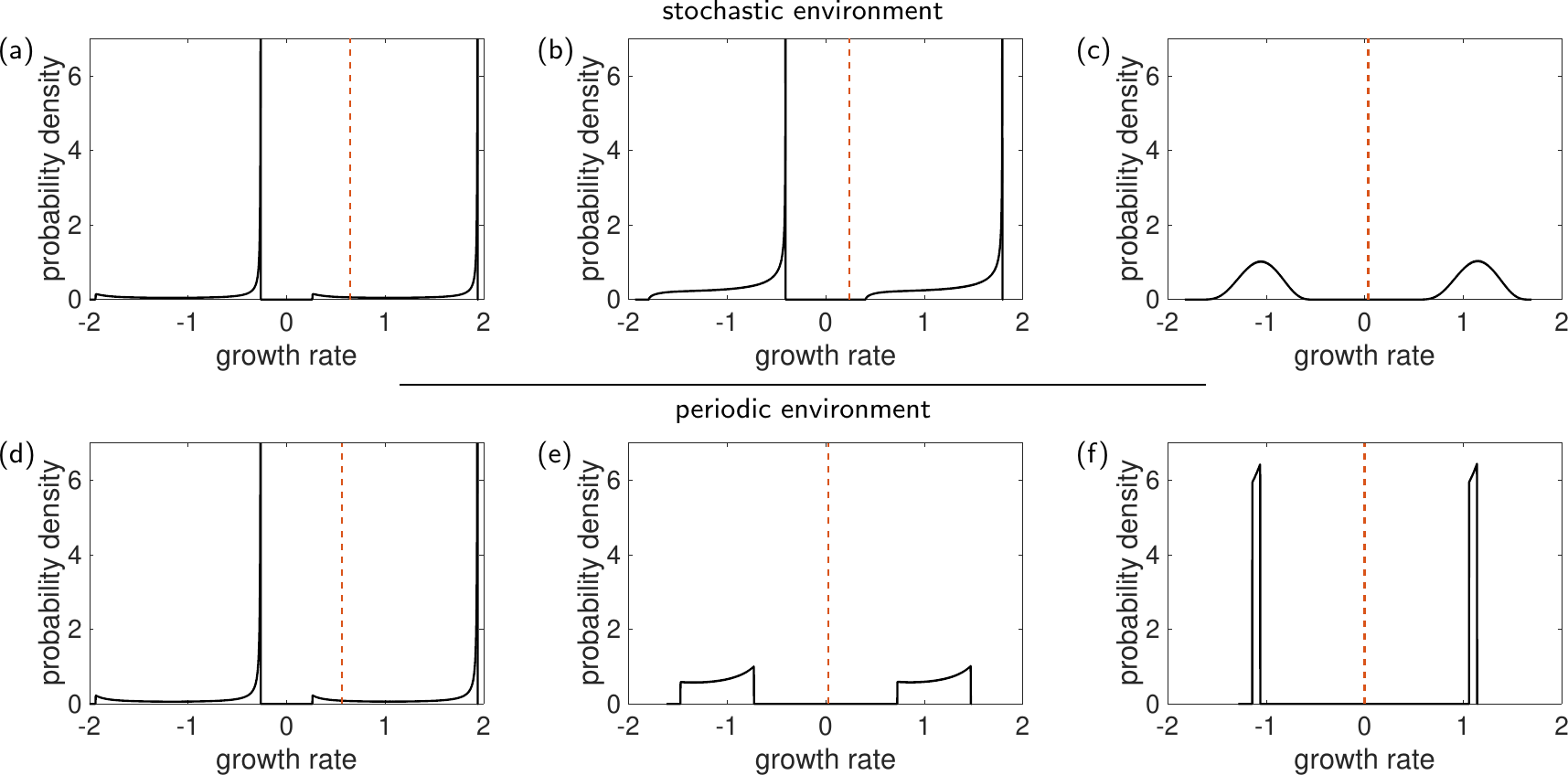}
\caption{Distribution of the instantaneous growth rate for the case of stochastic environmental dynamics [panels (a), (b), and (c)], and periodic environment [panels (d), (e), and (f)]. Environmental switching rates increase from the panels on the left to those on the right. Each distribution consists of two distinct components, corresponding to the two environments. The dashed line indicates the average growth rate.
Parameters are the same as in Fig.~\ref{fig:dist_phi}. }
\label{fig:dist_mu}
\end{figure*}

Panels (a)--(f) in Figure~\ref{fig:dist_phi} show the resulting stationary distributions for three different rates of environmental and phenotypic switching. In panels (a)--(c) we depict results for stochastic environmental dynamics, and in panels (d)--(f) the environment is periodic. The calculation of the stationary distribution for the periodic case is outlined in the Appendix~\ref{sec:periodic}. The orange line in each panel shows $\Pi^*_0(\phi)$ and the purple line shows $\Pi^*_1(\phi)$. As seen in the figure, the distributions are peaked near the two fixed points when environmental switching is slow [panels (a) and (d)]. At fast environmental switching rates, the system spends most of its time in the central region away from the fixed points [panels (c) and (f)].

The distributions in Eqs.~\eqref{eq:pdmp_dist} can be used to calculate the probability density of growth rates via Eq.~\eqref{eq:mu}. For the case of stochastic environments these are shown in Fig.~\ref{fig:dist_mu}(a--c). We proceed to study the fitness of a given phenotypic switching strategy (i.e., given switching rates $p$ and $q$). To do this we focus on the average growth rate 
\begin{equation}
{\rm{E}}\left(\mu\right) = P^*_0 {\rm{E}}\left(\mu\middle|0\right) + P^*_1 {\rm{E}}\left(\mu\middle|1\right),
\label{eq:mu_avg2}
\end{equation}
where we have
\begin{equation}
{\rm{E}}\left(\mu\middle|\sigma\right)= \mu_\sigma^A {\rm{E}}\left(\phi\middle|\sigma\right) + \mu_\sigma^B \left[ 1 - {\rm{E}}\left(\phi\middle|\sigma\right) \right],
\label{eq:mu_avg}
\end{equation}
and where $P^*_\sigma$ is the probability of finding the environment in state $\sigma$ in the stationary state. In the two-state system these are given by $P^*_0 = \lambda_0/(\lambda_1 + \lambda_0)$ and $P^*_1 = \lambda_1/(\lambda_1 + \lambda_0)$. The expectation value ${\rm{E}}\left(\phi\middle|\sigma\right)$ is given by
\begin{equation}
{\rm{E}}\left(\phi\middle|\sigma\right) = \int_{\phi_1^+}^{\phi_0^+} \frac{ \phi \Pi_\sigma^*(\phi)}{P^*_\sigma} \, \d \phi .
\label{eq:expect}
\end{equation}
Using Eqs.~\eqref{eq:pdmp_dist} we therefore have a closed integral equation for the average growth rate of the population. Evaluating Eq.~\eqref{eq:expect} numerically, and substituting into Eq.~\eqref{eq:mu_avg2} provides a very efficient way of calculating the average growth rate in terms of the model parameters. Results for the mean growth rate are indicated as dashed vertical lines in Fig.~\ref{fig:dist_mu}. 
The analysis below is for phenotypic switching rates, $p$ and $q$, which do not depend on the state of the environment. We note that the mathematical formalism applies to the case of environment-dependent switching rates as well; the calculation of the average growth rate can still be carried out if we replace $p$ and $q$ with environment-dependent rates $p_\sigma$ and $q_\sigma$. The dynamics are still reducible to a two-state PDMP for a single degree of freedom, and the stationary distribution can be obtained. Thus, the theory above provides an analytical formalism to investigate this case.
 
In the following section we use our solution to investigate the optimum switching strategy, i.e., the switching rates $p, q$ which allows the cells to best proliferate. Conversely, we also identify the optimum environmental switching rates, i.e., the switching rates $\lambda_0, \lambda_1$ which best inhibit the growth of cells.

\section{Results}
\label{sec:results}
In the context of our model, the biological strategy of bet-hedging refers to heterogeneity in the population to increase resilience against environmental changes, while at the same time maintaining the capability of growth. Specifically, we will quantify `temporal' heterogeneity below---the extent to which both phenotypes are present over the course of time, but not necessarily simultaneously. For example, a population may largely consist of only one phenotype at any one time, but as the environment changes state, both phenotypes may be expressed in turn. Viewed over time such a population would be heterogeneous.

In our model, in the absence of external stress (state $\sigma=0$), phenotype $A$ has a higher fitness. On the other hand, not having any individuals of type $B$ in the population is risky, as phenotype $A$ is more susceptible to the stresses in environment $\sigma=1$. The bet is hence hedged (the risk limited) by maintaining the capacity for a population of persister cells to establish itself should an external stress occur.
The analysis in Sec.~\ref{sec:analysis} provides the tools necessary to investigate the mechanics of bet-hedging. In particular, we have obtained an expression for the average growth rate, which can be evaluated numerically by a single integration. This significantly reduces the computing time required to analyse the dynamics, in contrast to direct Monte Carlo simulation of the individual-based model \cite{gillespie1977exact}. This allows us to explore a wide range of parameter space to investigate bet-hedging strategies.

\subsection{Optimal rate of phenotypic switching rate for stochastic environments}
\begin{figure*}
\includegraphics[width=0.94\textwidth]{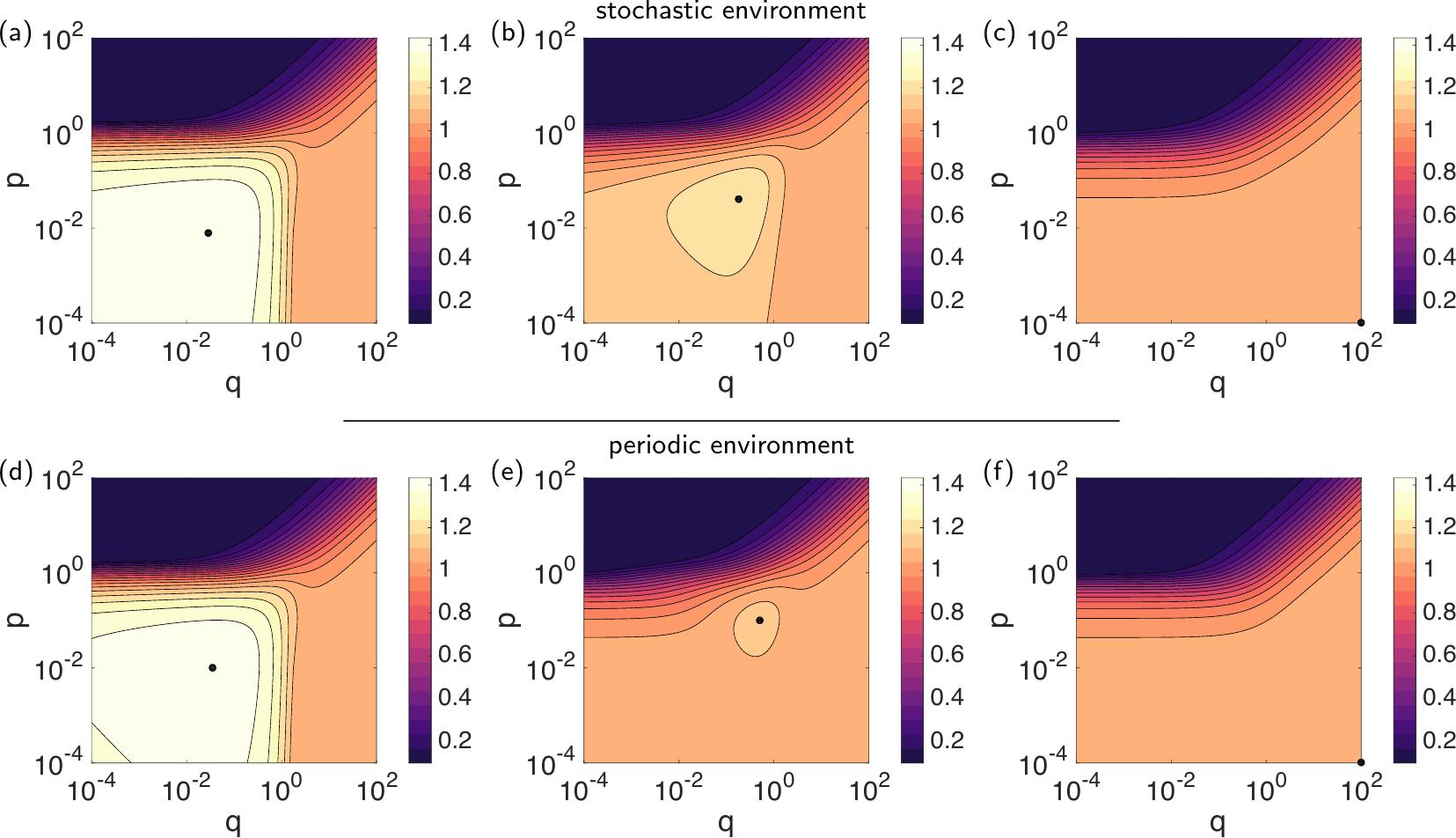}
\caption{Heatmap plot of the average growth rates for different values of the phenotypic switching rates $p$ and $q$. Panels (a), (b), and (c) show the case of a stochastic environment, and panels (d), (e), and (f) show that of a periodic environment. Environmental switching rates $\lambda_1$ and $\lambda_0$ increase from left [(a), (d)] to right [(c), (f)]. The black circle in each panel indicates the switching strategy which optimises the growth rate. When the environmental switching rates are slow, there is a non-trivial $p$ and $q$ maximising the growth rate. 
This means heterogeneity has a fitness advantage. When the environmental switching is fast, homogeneity is favoured. The stochastic case is found to have a higher growth rate than the periodic case for every value of $p$ and $q$.
Parameters set (a) (see Appendix~\ref{sec:parameters}) is used, along with: (a) and (d) $\lambda_1=0.010$, $\lambda_0=0.033$; (b) and (e) $\lambda_1=0.10$, $\lambda_0=0.33$; (c) and (f) $\lambda_1=1$, $\lambda_0=3.3$.}
\label{fig:heatmaps_klumpp}
\end{figure*}
We first study the dependence of the average growth rate on phenotypic switching rates, for given environmental switching dynamics. We consider different regimes of environmental dynamics, ranging from slow to fast, relative to the time scale of the growth of the population.

Figure~\ref{fig:heatmaps_klumpp} shows how the average growth rate depends on the phenotypic switching rates $p$ and $q$. The parameters $\mu_\sigma^A$ and $\mu_\sigma^B$, describing the growth rates of each phenotype, are fixed; panels (a), (b) and (c) correspond to regimes with increasingly fast environmental switching rates $\lambda_0$ and $\lambda_1$. When the environment switches relatively slowly, as in panels (a) and (b), there is a non-trivial phenotypic switching strategy which maximises the average growth rate, indicated by a black circle. When the environment switches very quickly, as in panel (c), the optimum strategy is found at the extremes of $p$ and $q$, indicating that it is best for the population to express only a single phenotype [phenotype $A$ in the case of Fig.~\ref{fig:heatmaps_klumpp}(c)]. These results are in agreement with previously reported work in similar systems \cite{wolf2005diversity,salathe2009evolution,gaal2010exact}. In other words, when environmental changes are slow, heterogeneity conveys a fitness advantage, whereas homogeneity is more beneficial in the face of fast environmental changes. 

\begin{figure*}
\includegraphics[width=0.95\textwidth]{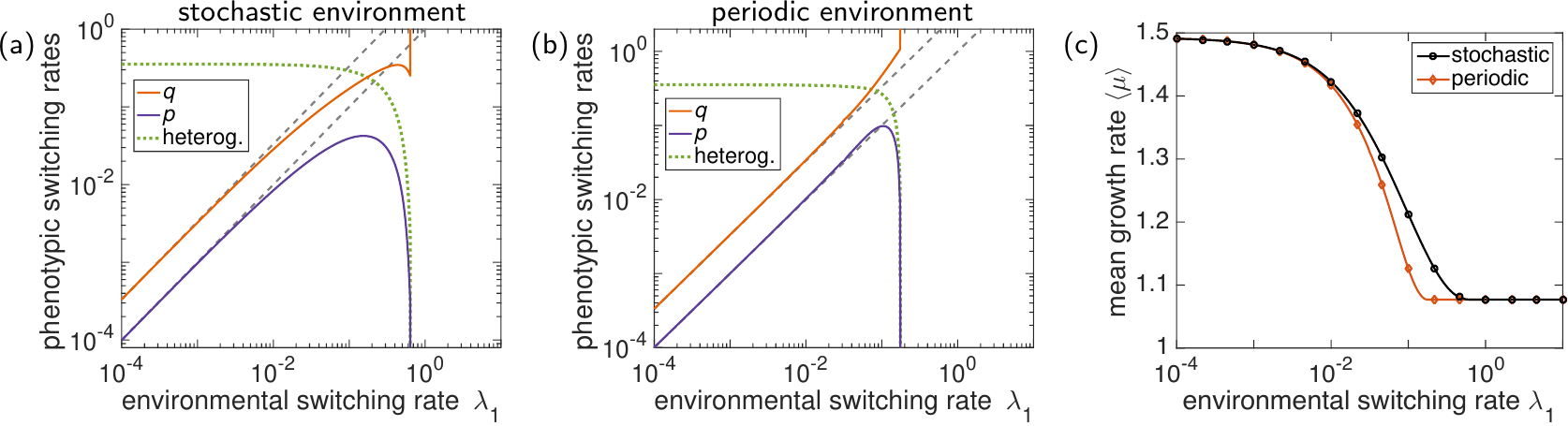}
\caption{(a) The optimum phenotypic response as a function of environmental switching rates for a stochastic environment.
The ratio of the environmental switching rates is fixed at $\lambda_0=\tfrac{10}{3}\lambda_1$. The vertical axis shows the phenotypic switching rates $p$ and $q$ which maximise the average growth rate. 
The grey, dashed lines show the environmental switching rates $\lambda_0$ and $\lambda_1$ for comparison.
When the environment switches slowly, the optimum strategy is given by matching the environmental rates: $p=\lambda_1, q=\lambda_0$. At faster environmental switching rates, the optimum strategy involves switching more slowly than the environment. At very fast environmental switching rates, the best strategy is to stay in one particular phenotype ($p\to 0$). (b) The optimum phenotypic as a function of environmental switching rates for a periodic environment. The periods the environments spends in state $0$ and $1$ have fixed durations, $1/\lambda_0$ and $1/\lambda_1$, respectively. At slow environmental switching rates, the optimum strategy is for the phenotypic switching rates to match the environment. At faster switching rates, the optimum strategy is to stay in a particular phenotype. For intermediate switching rates, phenotype switching rates slightly faster than the environment maximise the growth rate. The transition from heterogeneity to homogeneity occurs at slower environmental dynamics as in the stochastic case. The green dotted lines in panels (a) and (b) show the measure of temporal heterogeneity defined in the text. (c) Mean growth rate achieved by optimum choice of phenotypic switching rates $p$ and $q$ as we proportionally increase environmental switching rates. The markers indicate the results of Monte Carlo simulation of the PDMP and the periodic process. The mean growth rate in the case of stochastic environments is found to be universally larger than in periodic environments. Parameters $\mu_\sigma^A$ and $\mu_\sigma^B$ are as in the previous figures (set (a)).}
\label{fig:optimum_p_q_increasing_lambda}
\end{figure*}
This behaviour is further illustrated in Fig.~\ref{fig:optimum_p_q_increasing_lambda}(a). Here, we show how the optimum phenotypic switching rates vary as we increase the environmental switching rates $\lambda_0$ and $\lambda_1$, but keeping their ratio $\lambda_0/\lambda_1$ fixed, (i.e., the relative amount of time spent in each environment). We identify three regimes. When the environmental switching is slow, the optimum switching strategy is achieved when the phenotypic switching rates match the environmental rates: $p=\lambda_1$ and $q=\lambda_0$.

For environmental switching in an intermediate regime, the optimum strategy involves phenotypic switching rates which are much slower than the environmental switching rates. When the environment switches very quickly, the optimum strategy is not to switch between phenotypes at all, as discussed above. Instead, it is best for the cells to keep expressing whichever phenotype is better on average. That is, always express phenotype $A$ if $P^*_0\mu_0^A + P^*_1\mu_1^A<P^*_0\mu_0^B + P^*_1\mu_1^B$, and always phenotype $B$ otherwise. The growth rate is then $P^*_0\mu_0^A + P^*_1\mu_1^A$, or $P^*_0\mu_0^B + P^*_1\mu_1^B$, respectively. 

Our analytical findings for the regime of a slow environmental process are consistent with previous results by Thattai \emph{et al.}~\cite{thattai2004stochastic}, who studied both periodic and stochastic environments in numerical simulations. Results for periodically-switching environments can also be found in Refs.~\cite{wolf2005diversity, gaal2010exact}. Kussell and Leibler \cite{kussell2005phenotypic} report analytical results for randomly switching environments, using an approach based on maximising the average growth rate by minimising the transient time it takes a population to reach its environment-dependent quasi-stationary state.

Several quantitative measures of heterogeneity can in principle be considered. We focus on measuring `temporal' heterogeneity, i.e., the extent to which both phenotypes exist over the course of time. To this end, we use a variation of Simpson's diversity index \cite{simpson1949measurement}, and define heterogeneity as the probability that two cells selected at random from the population at two different times are of different phenotypes. Assuming that the individuals are sampled at two widely spread times, this probability is given by $2\left<\phi\right>\left(1-\left<\phi\right>\right)$. We have indicated this quantity in Fig. \ref{fig:optimum_p_q_increasing_lambda}(a) and (b). The data confirms that the population is heterogeneous for slow environmental switching, but homogeneous when the environmental dynamics are fast. 

Our calculation of the mean growth rates and composition of the population allows us to address a related question concerning the dependence of phenotypic heterogeneity and the optimal phenotypic switching rate on the ratio of the two environments $\lambda_0/\lambda_1$. We summarise our findings here, rather than present the details. In the regime of very slow environmental switching, the optimum strategy involves phenotypic switching for all values of the ratio $\lambda_0/\lambda_1$. This is because enough time passes in each environmental episode for the population to reach and benefit from its optimising quasi-stationary distribution. In the intermediate-switching regime, however, if a much longer time is spent on average in one environment than in the other it becomes optimal to express only the phenotype which performs best in this more frequent environment.

\subsection{Optimal rate of phenotypic switching rate for periodic environments} 
We perform a parallel study of the model with periodic environmental switching, described by Eq.~\eqref{eq:environ_periodic}. When the environment switches periodically and the infinite-population limit is taken, the dynamical system becomes deterministic. 
In the long run, the trajectory of $\phi(t)$ then converges to a limit cycle.
In environment $0$ phenotype $A$ is favoured; $\phi(t)$ increases until it reaches a `turning point' $\phi_\text{high}$ when the environment switches. Then, in environment $1$ phenotype $B$ is favoured; $\phi(t)$ decreases until turning point $\phi_\text{low}$, at which point the environment returns to state $0$.
It is straightforward to numerically evaluate $\phi_\text{high}$ and $\phi_\text{low}$.
As the trajectory tends to a limit cycle, we can compute the stationary distribution, i.e., the probability density of finding the variable $\phi(t)$ at a given point in the interval $[\phi_\text{low}, \phi_\text{high}]$ at a randomly chosen time. Further details of the mathematical method can be found in the Appendix~\ref{sec:periodic}. 

A comparison between the distributions of $\phi$ for stochastic and periodic switching environments is presented in Fig.~\ref{fig:dist_phi}. 
This figure shows that the domain in which the distribution of $\phi$ is non-zero is larger in the case of stochastic switching than for periodic environments. The effect is less pronounced for slow switching than for fast switching. When the environment switches stochastically, it is possible for the duration of each episode in a fixed state $\sigma$ to last longer than in the periodic case. As a result the trajectory of the PDMP can exceed the deterministic extremes $\phi_\text{low}$ and $\phi_\text{high}$ of the periodic case.

This difference has significant consequences for the growth rates and resulting phenotypic switching strategies. Figure~\ref{fig:dist_mu} shows the distribution of growth rates for both cases. As one can see, these distributions are broader for the case of stochastic environments than for periodic environmental changes, especially for intermediate and fast switching. 
Figure~\ref{fig:heatmaps_klumpp}(d)--(f) shows the numerically computed average growth rate in periodically switching environments as a function of phenotypic switching rates $p$ and $q$ (and for given $\lambda_0, \lambda_1$). The growth rates for each $p$ and $q$ is generally lower than that of the system with stochastic environmental dynamics with the same parameters.

Similarly to Figure ~\ref{fig:optimum_p_q_increasing_lambda}(a), Figure~\ref{fig:optimum_p_q_increasing_lambda}(b) shows the optimum phenotypic switching strategy, but for the case of a periodically switching environment. Superficially, the two figures look similar. In each we identify two extreme regimes of behaviour: (i) the limit of slow environmental switching where the optimum phenotypic strategy is to switch with the same rates as the environment, and (ii) the fast environmental regime where homogeneity is preferred. In the case of stochastic environmental dynamics, Figure \ref{fig:optimum_p_q_increasing_lambda}(a), an intermediate regime can be detected; this is however largely absent in the case of deterministic periodic environmental dynamics, where the transition between regimes (i) and (ii) is more abrupt [Figure \ref{fig:optimum_p_q_increasing_lambda}(b)]. A closer analysis reveals further differences between the cases of stochastic and periodic environments. For environmental dynamics at intermediate rates, i.e., between regimes (i) and (ii), the optimum phenotypic switching rates differ by as much as an order of magnitude between the stochastic and periodic cases. We also see that the stochastic case favours heterogeneity at much higher switching rates than the periodic case.

Figure~\ref{fig:optimum_p_q_increasing_lambda}(c) shows how the average growth rates compare in the cases of a Markovian environment and a periodically switching environment when the optimum phenotypic strategy is in effect. Stochastic switching produces a growth rate that is universally larger than in the periodic switching case. We found the same result for all tested parameters and for all values of phenotypic switching; similar results were previously also reported in Ref.~\cite{thattai2004stochastic}. We note that even small differences in the growth rate can lead to significant differences in population size over long times, as the growth is exponential. 

The explanation for this enhancement in growth rate is as follows: At any point in time, the instantaneous growth rate is given by Eq.~\eqref{eq:mu}.
The distribution of $\phi$ hence directly translates into a distribution of $\mu$, as illustrated in Figs.~\ref{fig:dist_phi} and~\ref{fig:dist_mu}. The difference between the growth rates for stochastic and periodic environments is most pronounced in the intermediate switching regime; the distributions of $\phi$ in this regime are shown Figs. \ref{fig:dist_phi}(b) and (e), for the cases of stochastic and periodic environments respectively. Figs.~\ref{fig:dist_mu}(b) and (e) show the corresponding distributions of growth rates. For periodic switching the distribution of $\phi$ is largely flat, with only minor peaks near the fixed points when the flow field in either environment is slow.
Conversely, stochastic environmental dynamics permit long episodes of time spent in a particular environment.
In such long episodes the variable $\phi$ will mostly reside near the relevant fixed point, generating the peaked behaviour seen in Fig.~\ref{fig:dist_phi}(b). Recalling that phenotype $A$ grows more quickly than $B$ in environment $0$ and vice versa in environment $1$, we conclude that the system is well-adapted to the environmental state at each of these peaks. This leads to an overall enhanced growth rate, relative to the periodic case.

The case of Markovian switching and the case of periodic switching are two extreme descriptions of the environmental process; we might expect a real-world environment to change with dynamics somewhere in-between \cite{thattai2004stochastic}. Our analysis suggests that, for an environment with dynamics between these two extremes, the average growth rate would fall somewhere in-between as well, and similarly for the optimum phenotypic switching rate.

\subsection{Optimal environmental switching strategy to inhibit population growth}
The environmental switching dynamics can be used to model a host immune response \cite{thattai2004stochastic,rainey2011evolutionary}. By switching between the two states, we assume the host aims to minimise the average growth rate of the population. 
In another application, the phenotypes can be wildtype bacteria and antibiotic-resistant strain, and the two environmental states represent the absence and presence of antibiotic treatment. The goal is then again to minimise growth, by varying the environmental protocol. Our analysis allows us to investigate the dependence of the growth rate on environmental switching rates $\lambda_0$ and $\lambda_1$. In this Section we investigate which environmental switching strategy best inhibits the growth of the population. The figures shown here are for stochastic environmental switching, however we found qualitatively similar results for periodic environments.

\begin{figure*}
\includegraphics[width=0.95\textwidth]{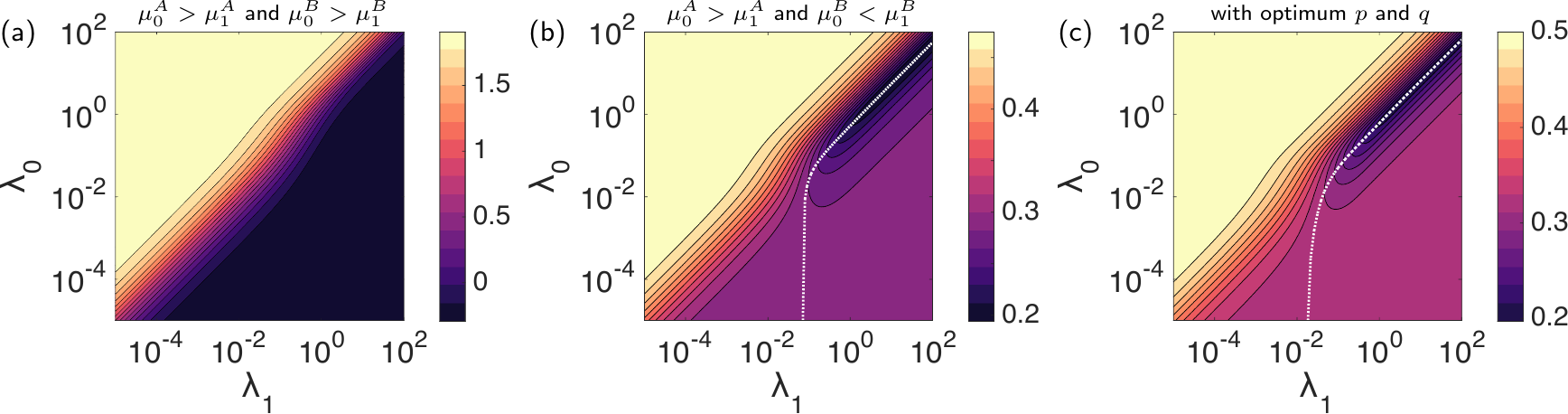}
\caption{(a) Heatmap plot of the average growth rate as a function of $\lambda_1$ and $\lambda_0$ using parameter set (a) as in the previous figures. The phenotypic switching rates are $p=0.1=q=0.1$. For these parameters the lowest growth rate is achieved when permanently in the stress state, $\lambda_0=0$. (b) The same quantity for parameter set (b), where $\mu^A_0>\mu^A_1$ and $\mu^B_0<\mu^B_1$. The phenotypic switching rates are fixed at $p=0.0275$, $q=0.0425$. The growth rate decreases as both switching rates are decreased in proportion. For a given $\lambda_0$ there is a non-trivial value of $\lambda_1$ which minimises the growth rate, shown here as a white line. It tend towards a straight line--a constant ratio which is given by Eq.~\eqref{eq:optimum_p0}: $\lambda_0/\lambda_1=0.557$ resulting in an average growth rate of $\left< \mu \right>=0.194$. (c) Heatmap plot for the {\em optimum} average growth rate as a function of the environmental switching rates: at each position on the graph the population of cells uses the optimum phenotypic switching strategy for the given environmental process. Parameters $\mu_\sigma^A$ and $\mu_\sigma^B$ as in panel (b). When the switching is sufficiently fast, there is a ratio of environmental switching rates which minimises the cell's best possible growth rate. This ratio is given by Eq.~\eqref{eq:optimum_ratio}.}
\label{fig:heatmap_k_k}
\end{figure*}

Figure~\ref{fig:heatmap_k_k}(a) and (b) show how the average growth rate changes if we vary the environmental switching rates for fixed phenotypic switching rates $p$ and $q$. Each panel represents a different set of parameters. The case shown in Figure~\ref{fig:heatmap_k_k}(a) is a representation of bacteria in which both phenotypes are disadvantaged by the presence of an antibacterial agent, but in which the persister phenotype is affected less by the external stress than the normal phenotype. Mathematically one has $\mu_0^A>\mu_1^A$, $\mu_0^B>\mu_1^B$, $\mu_0^A>\mu_0^B$, and $\mu_1^B>\mu_1^A$. Specifically, we use the parameters set used in previous studies \cite{patra2015emergence}, and we will refer to this as parameter set (a) in the text which follows; the numerical values can be found in the Appendix~\ref{sec:parameters}. Unsurprisingly, the average growth rate is minimised when the environment stays in environmental state $1$ (i.e., $\lambda_0=0$), which can be seen as representing the antibiotic state. Another trend shown by our results is that proportionally increasing both environmental switching rates universally decreases the average growth rate. That is, proportionally faster switching rates better inhibit the growth of a population. This result has been found for all tested parameters, suggesting that the best way apply antibiotics may involve many short periods of antibiotic treatment.

The parameters used in Fig.~\ref{fig:heatmap_k_k}(b) differ from those in Fig.~\ref{fig:heatmap_k_k}(a) in that here $\mu^A_0>\mu^A_1$ and $\mu^B_0<\mu^B_1$. We use the parameters of Ref.~\cite{belete2015optimality}, and will refer to this as parameter set (b) in the text which follows; numerical values are again given in Appendix~\ref{sec:parameters}. Here, phenotype $A$ performs better in environment $0$ than in environment $1$, and phenotype $B$ performs better in environment $1$ in environment $0$. In this case, the optimum environmental strategy involves non-trivial switching. The white line in panel Fig.~\ref{fig:heatmap_k_k}(b) indicates the optimum choice of $\lambda_1$, for given $\lambda_0$. When both $\lambda_0$ and $\lambda_1$ are large growth is minimised for a non-trivial ratio, $\lambda_1/\lambda_0$.
 
The limit of infinitely fast environmental switching can be characterised analytically. In this limit, the dynamics can be modelled using ordinary differential equations, using effective mean birth and death rates \cite{haseltine2002approximate, rao2003stochastic}. Minimising the growth rate in this limit leads to an equation describing the optimum environmental strategy to inhibit the growth of the population. The minimal growth rate is found when $P^*_0$, the probability of being in environmental state $0$, is given by
\begin{equation}
P^*_0=\frac{-\Delta_1+p-q}{\Delta_0-\Delta_1} - \left(\frac{-pq}{(\mu_0^A-\mu_1^A)(\mu_0^B-\mu_1^B)} \right)^{1/2} .
\label{eq:optimum_p0}
\end{equation}
The argument of the square root indicates a necessary condition (up to relabelling) for the existence of this optimum, $\mu_0^A>\mu_1^A$ and $\mu_0^B<\mu_1^B$. This explains why we have no non-trivial minimum in the fast-switching limit for the parameters used in Fig.~\ref{fig:heatmap_k_k}(a), but there is one for the parameters used in Fig.~\ref{fig:heatmap_k_k}(b). The analytical optimum given by Eq.~\eqref{eq:optimum_p0} agrees with the results shown in Fig.~\ref{fig:heatmap_k_k}(b).

\subsection{Co-evolution of host and pathogen}
The interplay between the growth-maximising strategy of the cells and the growth-minimising strategy of the environment provides an interesting scenario. Given a set of environmental switching rates, there can be phenotypic switching rates which maximise the average growth rate. Conversely, for given phenotypic switching rates, there can be non-trivial environmental switching rates which minimise the growth rate. This suggests the possibility of a combined set of rates, from which the growth rate cannot be increased by unilaterally changing $p$ and $q$ (the `strategy' of the population), and from which the growth rate cannot be decreased by unilaterally changing $\lambda_1$ and $\lambda_0$ (the strategy of the hosting environment). From a game-theoretic perspective, neither player of the game (the population of cells and the immune system) can improve their performance by changing their strategy. This can be viewed as a Nash equilibrium \cite{nash1951non}.

For a given set of model parameters, $\mu_\sigma^A$, $\mu_\sigma^B$ and given environmental switching rates, optimal phenotypic switching rates can be found by evaluating Eq.~\eqref{eq:mu_avg2} for different values of $p$ and $q$, and then identifying the maximum. This maximum can then be studied as a function of $\lambda_0$ and $\lambda_1$. For parameter set (a) the evolutionarily optimising strategy is found to be trivial: the environment stays in the stress state and the cell expresses the persister phenotype only, leading to an overall growth rate $\mu^B_1$.
 
For parameter set (b), however, the results are less trivial, since each phenotypes prefers a different environment. Figure~\ref{fig:heatmap_k_k}(c) shows the average growth rate given optimal phenotypic switching in this case. This figure was constructed as follows: for each combination of $\lambda_0$ and $\lambda_1$ we found the values of $p$ and $q$ maximising the growth rate. The value of this maximum is then plotted in the figure. For each value of $\lambda_1$ the white line in the figure depicts the value of $\lambda_0$, which minimises this maximum achievable growth rate; from the point of view of the hosting environment this line shows the best choice of $\lambda_0$ for each $\lambda_1$. As before, the growth rate decreases as the environmental switching rates are increased proportionally. Hence, the environment's best strategy is to switch states quickly, and with a choice of $\lambda_0/\lambda_1$ so that switching rates fall on the white line in Figure~\ref{fig:heatmap_k_k}(c). 

As discussed in earlier sections the maximising phenotypic strategy in the limit of fast environments leads to homogeneity, i.e., it is best for the population to only express the phenotype does better in a (weighted) average over both environmental states. This gives a growth rate which is the greater of $P^*_0\mu_0^A + P^*_1\mu_1^A$ and $P^*_0\mu_0^B + P^*_1\mu_1^B$.
Inspecting these expressions, one finds that changing the ratio of $\lambda_0$ and $\lambda_1$ increases one of the quantities and decreases the other.
It follows that the greater of the two growth rates is minimised when both growth rates are equal. This the case when
\begin{equation}
\frac{\lambda_0}{\lambda_1}=-\frac{\Delta_1}{\Delta_0}.
\label{eq:optimum_ratio}
\end{equation}
This gives us the proportion of each environmental state which minimises the average growth rate in the limit of fast environmental switching, given an optimally switching population. This minimum optimum growth rate is given by
\begin{equation}
\left< \mu \right> = \frac{\mu_1^B\mu_0^A - \mu_0^B\mu_1^A}{\mu_0^A+\mu_1^B-\mu_0^B-\mu_1^A }.
\end{equation}
The phenotypic switching strategy of the cell cannot change to increase the growth rate. Similarly, if the environment assumes that the cell population will optimise its phenotypic switching strategy, the environment cannot move to minimise the growth rate. We remark that the final rate is independent of the phenotypic switching rates $p$ and $q$, i.e., the environment's minimisation of the growth rate removes the effect of the phenotypic switching strategy altogether.

The analytical result from Eq.~\eqref{eq:optimum_ratio} agrees with the ratio given defined by the fast-switching asymptote of the white line in Fig.~\ref{fig:heatmap_k_k}(c). The minimum growth rate in this case is $0.197$, which is significantly less than if the environmental states did not switch, which would result in growth rates $0.5$ or $0.325$, respectively, for the two environmental states.

\section{Conclusion}\label{sec:conclusion}
In conclusion we have studied a stylised model of phenotypic switching strategies in the face of changing environments. We have focused on the role of time scales of phenotypic and environmental switching for the growth of the population, and our analysis addresses in particular the case of stochastically switching environments. 

In contrast to some existing work, our analysis starts from an explicit individual-based model of the population dynamics, defined by birth events, death events, and stochastic phenotypic switching. Our analysis then proceeds through the formulation of a piecewise-deterministic Markov process. This allows us to derive closed-form solutions for the resulting growth rates of the population for general environmental and phenotypic switching rates. Our work complements existing literature which, for mathematical convenience, has often concentrated on strictly periodic environmental dynamics. We systematically compare the cases of periodic versus stochastically switching external conditions. Our main results can be summarised as follows: (i) Using our theory, we demonstrated that the optimal phenotypic switching rates are equal to the rates of the environmental process ($p=\lambda_1, q=\lambda_0$) for slow stochastic environmental dynamics. This result was previously reported in simulations \cite{thattai2004stochastic, kussell2005phenotypic}. For environmental dynamics in an intermediate regime the optimal switching rates are markedly lower than those of the environment.
(ii) The optimal bet-hedging strategy of the bacterial population favours heterogeneity (both phenotypes present) in slow environments, but is replaced by a homogeneous strategy (one phenotype only) for fast switching. The transition between these regimes is sudden as the switching rates of the environmental process are increased.
We find that stochastic environments favour heterogeneity over a larger range of environmental dynamics than a strictly periodic environment.
(iii) Instantaneous growth rates are universally higher in the case of stochastic environments than in the periodic case. Our analysis shows that this due to the possibility to spend long periods of time in either environment when the environmental process is stochastic---a possibility that is absent for strictly periodic environments. (iv) The combined system of population and environment can be interpreted as the interaction between a hosting environment and a pathogen. The host tries to control environmental switching to minimise growth of the pathogen. The pathogen, on the other hand, attempts to maximise its growth rate by phenotypic switching with optimised rates. Our analysis shows that mutual best-response scenarios can be identified, in which neither the host nor the pathogen can improve by unilateral changes of their strategy. This is akin to the concept of Nash equilibria in game theory.

In this paper, we developed a mathematical framework for the analysis of phenotypic switching in stochastic environments. More broadly, our work is applicable to ecological models of competition in dynamic random external conditions.
Such problems are widespread in theoretical ecology, in predator-prey models effectively random environments could for example account for external shocks such as earthquakes and epidemics \cite{luo2007stochastic,zhu2009hybrid,bao2011competitive}. Our work can provide insights into the mechanics of such problems, and as a key contribution, the analytical computation of average growth rates allows one to dispense with costly Monte Carlo simulations. 

We stress that the model we have focused on is stylised and could naturally be extended in many different ways to describe more realistic and complicated features. This may include models in which the dynamics of the environment are coupled to the state of the population of microbial species \cite{visco2010switching}.
Further complications also occur when there are more than two environmental states. We have chosen the stylised scenario of a binary environment as it represents situations in which an external stress is either absent or present. A model with two environmental states and two phenotypes, each doing best in one of the two environments, is a minimal, but non-trivial baseline. While it is unsurprising that much of the existing literature has focused on this case, models with more environmental states have for example been studied in \cite{kussell2005phenotypic}. To obtain interesting scenarios it is then also necessary to introduce multiple phenotypes. While a detailed extension of our analysis to such cases is pending we expect many of the results to be qualitatively similar in this case. 
Several recent papers \cite{assaf2013extrinsic,roberts2015dynamics,assaf2013cooperation} consider the role of a continuous environment, and it would be interesting and challenging to establish the exact relation between these two approaches. We highlight this as a potential area for future research. We note in particular that environmental processes based on stochastic differential equations may lead to scenarios in which the environment spends significant time in intermediate states, a possibility that is excluded by construction in our model. It is not clear if, when and how these differences affect the response of the population.
The phenotypic switching rates themselves are treated as static in our model; this is an approximation as well, bacteria have been demonstrated to sense and adapt to external conditions \cite{dorr2009sos,dorr2010ciprofloxacin}. Incorporating this type of SOS response is another line of future work to make models of phenotypic switching more realistic. The formalism we have presented can be applied to study such cases.

\section*{Acknowledgements}
We acknowledge the Engineering and Physical Sciences Research Council (EPSRC, UK) for funding in form of a PhD studentship to PGH, and grant No. EP/K037145/1 supporting YTL and TG. YTL thanks the Center for Nonlinear Studies for supporting continuation of this work. We thank Stefan Kumpp for discussions.

\section*{Author contributions}
All three authors contributed to conceiving and designing the study. PGH and YTL carried out the analytical work and numerical simulations. All authors wrote and reviewed the paper.

\section*{Competing interests}
The authors declare no competing interests.

\appendix

\section{Model parameters}
\label{sec:parameters}
We consider the model for two different sets of parameters, $\mu_\sigma^{A,B}$. Parameter set (a) was previously used by Patra and Klumpp \cite{patra2015emergence}, and is shown in Table~\ref{table:parameters}(a). In this case, both phenotypes $A$ and $B$ perform better in environment $0$ (no external stress) than in environment $1$ (with external stress). The persister phenotype $B$ grows more slowly state $0$, but also decays less quickly in state $1$. This model is a classic representation of bacteria in which both phenotypes are disadvantaged by the presence of an antibacterial agent, but in which the persister phenotype is affected less by the external stress than the normal phenotype. Mathematically one has $\mu_0^A>\mu_1^A$, $\mu_0^B>\mu_1^B$, $\mu_0^A>\mu_0^B$, and $\mu_1^B>\mu_1^A$.

The second set of parameters was previously used by Belete and Bal\'{a}zsi \cite{belete2015optimality}, see Table~\ref{table:parameters}(b). These parameters have different properties: phenotype $B$ now performs better in environment $1$ than it does in environment $0$; i.e., $\mu_0^A>\mu_1^A$, $\mu_0^B<\mu_1^B$, $\mu_0^A>\mu_0^B$, and $\mu_1^B>\mu_1^A$.

\begin{table}[h!]
\centering
 (a)~\begin{tabular}{| l | r |}
 \hline
 $\mu_0^A$ & $2.0 \text{ h}^{-1}$\\ \hline 
 $\mu_0^B$ & $0.2 \text{ h}^{-1}$\\ \hline
 $\mu_1^A$ & $-2.0 \text{ h}^{-1}$\\ \hline
 $\mu_1^B$ & $-0.2 \text{ h}^{-1}$\\ \hline
 \end{tabular}
 \hspace{5mm}(b)~\begin{tabular}{| l | r |}
 \hline
 $\mu_0^A$ & $0.5000 \text{ h}^{-1}$\\ \hline
 $\mu_0^B$ & $0.0001 \text{ h}^{-1}$\\ \hline
 $\mu_1^A$ & $0.0001 \text{ h}^{-1}$\\ \hline
 $\mu_1^B$ & $0.3250 \text{ h}^{-1}$\\ \hline
 \end{tabular}\caption{The two parameters sets: (a) from Ref.~\cite{patra2015emergence}, and (b) from Ref.~\cite{belete2015optimality}.}
 \label{table:parameters}
\end{table}

\section{Derivation of diffusive process with Markovian switching}
\label{sec:derivation}
\subsection{Kramers--Moyal expansion}
The model individual-based describes the evolution of three random processes: the number of individuals with phenotype $A$ ($a_t$), the number of individuals with phenotypes $B$ ($b_t$), and the state of the environment ($\sigma_t\in\{0,1\}$). The master equation describes the time evolution of the probability distribution of these random processes, and is given by
\begin{eqnarray}
\begin{aligned}
\frac{\d}{\d t}P_{a,b,\sigma}(t)=
{}&\left( \mathcal{E}_{a}^{-1}-1 \right) \alpha_\sigma a P_{a,b,\sigma}(t)\\
+&\left( \mathcal{E}_{b}^{-1}-1 \right) \beta_\sigma b P_{a,b,\sigma}(t)\\
+&\left( \mathcal{E}_{a}^{+1}-1 \right) \gamma_\sigma a P_{a,b,\sigma}(t)\\
+&\left( \mathcal{E}_{b}^{+1}-1 \right) \delta_\sigma b P_{a,b,\sigma}(t)\\
+&\left( \mathcal{E}_{a}^{+1}\mathcal{E}_{b}^{-1}-1 \right) p a P_{a,b,\sigma}(t)\\
+&\left( \mathcal{E}_{a}^{-1}\mathcal{E}_{b}^{+1}-1 \right) q b P_{a,b,\sigma}(t)\\
+&\lambda_\sigma P_{a,b,1-\sigma}(t) - \lambda_{1-\sigma} P_{a,b,\sigma}(t)
\label{eq:master_equation}
\end{aligned}
\end{eqnarray}
where $P_{a,b,\sigma}(t)$ is the probability of random processes $(a_t,b_t,\sigma_t)$ having values $(a,b,\sigma)$ at time $t$.
 
For compactness, we suppress the explicit time dependence of the probability distribution. The notation $\mathcal{E}_{a}^{\pm}$ and $\mathcal{E}_{a}^{\pm}$ represents the step operators,
\begin{equation}\begin{aligned}
\mathcal{E}_{a}^{\pm}f(a,b)=&f\left(a\pm1,b\right), \nonumber \\
\mathcal{E}_{b}^{\pm}f(a,b)=&f\left(a,b\pm1\right),
\end{aligned}\end{equation}
where $f(a,b)$ is a generic function of $a$ and $b$.

We proceed a to approximate the master equation by means of a Kramers--Moyal expansion \cite{van1992stochastic, gander2007stochastic}, i.e., we replace the step operators by the first two non-trivial terms in their Taylor expansion. The variables $a$ and $b$ become continuous during this process, which is valid in the limit of large populations. Collecting terms up to order $a^{-2}$ or $b^{-2}$ and maintaining the discreteness of the environmental switching we find that the probability density $\Pi_{\sigma}(a,b)$ is governed by the following equation
\begin{eqnarray}
\begin{aligned}
\partial_t\Pi_{\sigma}(a,b)=
-&\partial_a \left( \mu_\sigma^A a -pa+qb \right)\Pi_{\sigma}(a,b)\\
-&\partial_b \left( \mu_\sigma^B b +pa -qb \right)\Pi_{\sigma}(a,b)\\
+&\tfrac{1}{2}\partial_a^2 \left( \alpha_\sigma a + \gamma_\sigma a +p a + qb \right)\Pi_{\sigma}(a,b)\\
+&\tfrac{1}{2}\partial_b^2 \left( \beta_\sigma b + \delta_\sigma b +p a + qb \right)\Pi_{\sigma}(a,b)\\
-&\partial_a\partial_b \left( pa+qb\right)\Pi_{\sigma}(a,b)\\
+&\lambda_\sigma \Pi_{\sigma'}(a,b) - \lambda_{\sigma'} \Pi_{\sigma}(a,b).
\label{eq:FPE_a_b}
\end{aligned}
\end{eqnarray}
The first two terms on the right-hand side are the drift terms as found in the Fokker--Planck equation \cite{risken1984fokker}, describing the flow of probability along the $a$ and $b$ directions, the next three terms describe diffusion, and the final two terms represent the random switching between the two environmental states. This Fokker--Planck equation has the diffusion matrix
\begin{align*}
D_\sigma=\left( \begin{array}{cc}
a\left(\alpha_\sigma+\gamma_\sigma\right)+pa+qb
& -pa-qb\\
-pa-qb
& b\left(\beta_\sigma+\delta_\sigma\right)+pa+qb
\end{array} \right).\numberthis
\end{align*}
The process described by the Fokker--Planck equation can be equivalently formulated as a stochastic differential equation, 
\begin{subequations}\begin{align}
\begin{split}d a_t=&\left(\mu_\sigma^A a_t -pa+qb\right)\d t \\&+ {B_\sigma}_{11}(a_t,b_t) \d W_t^{(1)} + {B_\sigma}_{12}(a_t,b_t) \d W_t^{(2)},\end{split}\\
\begin{split}\d b_t=&\left(\mu_\sigma^B b_t +pa-qb\right)\d t \\&+ {B_\sigma}_{21}(a_t,b_t) \d W_t^{(1)} +{B_\sigma}_{22}(a_t,b_t) \d W_t^{(2)},\end{split}
\end{align}\label{eq:appendix_sde_a_b}\end{subequations}
where the matrix $B_\sigma$ fullfills $B_\sigma^2=D_\sigma$. Specifically, we have
\begin{equation}
B_\sigma(a,b)=\frac{1}{r}\left( \begin{array}{cc}
a\left(\alpha_\sigma+\gamma_\sigma\right)+pa+qb+s
& -pa-qb,\\
-pa-qb
& b\left(\beta_\sigma+\delta_\sigma\right)+pa+qb+s,
\end{array} \right)
\label{eq:B}
\end{equation}
where we have introduced the shorthand
\begin{subequations}\begin{align*}
 s =&\big[ab\left(\alpha_\sigma+ \gamma_\sigma\right)\left(\beta_\sigma+\delta_\sigma\right)+a\left(\alpha_\sigma+ \gamma_\sigma\right)(pa+qb)
+b\left(\beta_\sigma+\delta_\sigma\right)(pa+qb)\big]^{\frac{1}{2}},\numberthis\\
r =&\left[a\left(\alpha_\sigma+ \gamma_\sigma\right)+b\left(\beta_\sigma+\delta_\sigma\right)+2pa+2qb+2s\right]^\frac{1}{2}.\numberthis
\end{align*}\end{subequations} 
The process $\sigma_t$ remains discrete and is described by the master equation
 \begin{align}
\frac{\d}{\d t}P_{\sigma}=&\lambda_\sigma P_{1-\sigma} - \lambda_{1-\sigma} P_{\sigma}.
\end{align}
\subsection{Transformation of coordinates}
We are able to make analytical progress by changing to relative coordinates. We introduce the random processes $n_t=a_t+b_t$ and $\phi_t=a_t/n_t$, describing the total population and fraction expressing phenotype A, respectively. Using Eq.~\eqref{eq:appendix_sde_a_b} and applying the rules of Ito calculus we find
\begin{subequations}\begin{align}
\d n_t=&n_t\left(\mu_\sigma^A \phi_t + \mu_\sigma^B (1-\phi_t)\right)\d t + n_t^{\frac{1}{2}}{C_\sigma}_{11}(\phi_t) \d W_t^{(1)} + n_t^{\frac{1}{2}}{C_\sigma}_{12}(\phi_t) \d W_t^{(2)}+\mathcal{O}(n^0),\\
\d \phi_t=&\left[\Delta_\sigma \phi_t(1-\phi_t) -p\phi_t+q(1-\phi_t) \right]\d t + n_t^{-\frac{1}{2}}{C_\sigma}_{21}(\phi_t) \d W_t^{(1)} + n_t^{-\frac{1}{2}}{C_\sigma}_{22}(\phi_t) \d W_t^{(2)}+\mathcal{O}(n^{-1}),
\end{align}\end{subequations}
where
\begin{equation}
C_\sigma(\phi)=\frac{1}{r'}\left( \begin{array}{cc}
\phi\left(\alpha_\sigma+\gamma_\sigma\right)+s'~~
& (1-\phi)\left(\beta_\sigma+\delta_\sigma\right)+s'\\
(1-\phi)\left[\phi\left(\alpha_\sigma+\gamma_\sigma\right)+s'\right]+p\phi+q(1-\phi)~~
&-\phi\left[(1-\phi)\left(\beta_\sigma+\delta_\sigma\right)+s'\right]-p\phi-q(1-\phi)\\
\end{array} \right)\label{eq:C}
\end{equation}
and
\begin{subequations}\begin{align}
s'=&\sqrt{\phi(1-\phi)\left(\alpha_\sigma+ \gamma_\sigma\right)\left(\beta_\sigma+\delta_\sigma\right)+\phi\left(\alpha_\sigma+ \gamma_\sigma\right)[p\phi-q(1-\phi)]+(1-\phi)\left(\beta_\sigma+\delta_\sigma\right)[p\phi+q(1-\phi)]},\\
r'=&\sqrt{\phi\left(\alpha_\sigma+ \gamma_\sigma\right)+(1-\phi)\left(\beta_\sigma+\delta_\sigma\right)+2p\phi+2q(1-\phi)+2s'}.
\end{align}\end{subequations}

\section{Growth rate in periodic environments}
\label{sec:periodic}
In the periodic case the environment is described by Eq.~\eqref{eq:environ_periodic}. Neglecting intrinsic fluctuations the entire dynamical system becomes deterministic, and the evolution of $\phi$ is described by the ordinary differential equation
\begin{equation}
\frac{\d \phi}{\d t} = \Delta_{\sigma(t)} \phi (1-\phi)-p\phi+q(1-\phi). \label{eq:PDMP_phi_periodic}
\end{equation}
In the long run, any trajectory of $\phi(t)$ converges to a limit cycle: in environmental state $0$ phenotype $A$ is favoured, and so $\phi(t)$ increases to a `turning point' $\phi_\text{high}$ at which point the environment switches. In environmental state $1$ phenotype $B$ is favoured, so $\phi(t)$ decreases until another turning point $\phi_\text{low}$ at which point the environment switches again. The turning points $\phi_\text{low}$ and $\phi_\text{high}$ can be found by numerical integration of Eq.~\eqref{eq:PDMP_phi_periodic}.

As the trajectory tends to a limit cycle, we can compute the resulting distribution for $\phi$, i.e., the fraction of time $\phi(t)$ spends in a given interval during a cycle. This distribution is limited to the on the domain $[\phi_\text{low}, \phi_\text{high}]$. We let $T=1/\lambda_1 + 1/\lambda_0$ be the period of the limit cycle. Since the time $\d t$ spent in a specific range $\d \phi$ is given by Eq.~\eqref{eq:PDMP_phi_periodic}, we find
\begin{equation}
\Pi^*_\sigma(\phi)=\frac{1}{T}\frac{1}{\left| \Delta_\sigma \phi (1-\phi)-p\phi+q(1-\phi)\right|},
\numberthis
\end{equation}
for $\phi\in[\phi_\text{low},\phi_\text{high}]$ . These distributions are analogous to the stationary distributions when the environments switch stochastically. The remaining analysis the same as Eq.~\eqref{eq:mu_avg} onwards.

\end{document}